\documentclass[letter,scriptaddress,apl,showkeys,superscriptaddress,preprint]{revtex4}

\usepackage{amsmath}
\usepackage{makeidx}
\usepackage{amsfonts}
\usepackage[ansinew]{inputenc}
\usepackage[usenames,dvipsnames]{pstricks}
\usepackage{subfigure}
\usepackage{graphicx}
\usepackage{dcolumn}
\usepackage{bm}
\usepackage{float}
\restylefloat{figure}
\usepackage{epsfig}
\usepackage{pst-grad} 
\usepackage{pst-plot} 

\makeindex
\begin{document}

\title{Ultrafast electrical control of a resonantly driven single photon source\\}

\author{Y. Cao}
\affiliation{Toshiba Research Europe Limited, Cambridge Research Laboratory, 208 Science Park, Milton Road, Cambridge, CB4 0GZ, United Kingdom}
\affiliation{Controlled Quantum Dynamics Group, Imperial College London, London, SW7 2AZ, United Kingdom}
\author{A. J. Bennett}
\email{anthony.bennett@crl.toshiba.co.uk}
\affiliation{Toshiba Research Europe Limited, Cambridge Research Laboratory, 208 Science Park, Milton Road, Cambridge, CB4 0GZ, United Kingdom}
\author{D. J. P. Ellis}
\affiliation{Toshiba Research Europe Limited, Cambridge Research Laboratory, 208 Science Park, Milton Road, Cambridge, CB4 0GZ, United Kingdom}
\author{I. Farrer}
\affiliation{Cavendish Laboratory, University of Cambridge, JJ Thomson Avenue, Cambridge, CB3 0HE, United Kingdom}
\author{D. A. Ritchie}
\affiliation{Cavendish Laboratory, University of Cambridge, JJ Thomson Avenue, Cambridge, CB3 0HE, United Kingdom}
\author{A. J. Shields}
\affiliation{Toshiba Research Europe Limited, Cambridge Research Laboratory, 208 Science Park, Milton Road, Cambridge, CB4 0GZ, United Kingdom}

\date{\today}

\begin{abstract}
We demonstrate generation of a pulsed stream of electrically triggered single photons in resonance fluorescence, by applying high frequency electrical pulses to a single quantum dot in a p-i-n diode under resonant laser excitation. Single photon emission was verified, with the probability of multiple photon emission reduced to $2.8\%$. We show that despite the presence of charge noise in the emission spectrum of the dot, resonant excitation acts as a ``filter'' to generate narrow bandwidth photons.
\end{abstract}

\pacs{Valid PACS appear here}
\maketitle
A high quality, non-classical source of on-demand single photons represents the essential common denominator for a variety of applications of quantum technologies, such as quantum key distribution \cite{Scarani2009}, long distance quantum communication \cite{Yin2012} and building scalable quantum computing architectures \cite{Lindner2009}. Substantial progress has been made towards implementing electrically \cite{Reischle2010,Faraon2010} and optically driven \cite{NickVamivakas2009,Ulhaq2012,Muller2007,Yilmaz2010,Nguyen2011,Greve2013} single photon sources as well as devices that enable the manipulation of quantum bits\cite{Laurent2005,Fallahi2010,Gao2012,Godden2012} based around semiconductor quantum dots. Quantum dots embedded in electrical devices enable control over properties such as the emission energy \cite{Finley2004}, the g-factor \cite{Jovanov2011} and the fine structure splitting \cite{Marcet2010}. However, all quantum dots suffer from noise, caused by trapping and releasing of charges in localization centres near to the quantum dot \cite{Kuhlmann2013a} or through electrical contacts. These sources of noise induce fluctuation in the charge environment of the exciton, which can lead to undesirable effects such as reduced single photon indistinguishability \cite{Patel2010} and spectral wandering \cite{Robinson2000}. To avoid these effects while maintaining the tunable properties, a combination of electrical and optical controls maybe required. In this study, we present an electrically tunable quantum dot in a p-i-n type device, driven by resonant $s$-shell excitation and show that the emitted photons are impervious to charge induced decoherence. By applying an ultra-high frequency electrical pulse train, we demonstrate on-demand single photon emission from this device, under continuous wave laser excitation.

Our single photon generation scheme employs self-assembled InAs/GaAs quantum dots embedded inside a p-i-n diode. Quantum Dots are located at the center of a $10\:nm$ GaAs layer, with a $70\:nm$ AlGaAs super-lattice on both sides. This structure enabled precise tuning of emission energies in a range of up to $25\:meV$ \cite{Bennett2010} through the quantum confined Stark effect. Emission energies of the exciton ($X$), biexciton ($XX$) and charge excitons ($X^{\pm}$) were controlled by applying a bias voltage between the bottom ohmic contact and a $Ti/Au$ top-surface electrode. The binding energies $E_{B}(X/X^{-})$ measured at $0.2\:V$, defined as $E_{B}(X)=E_{X}-E_{XX}$ and $E_{B}(X^{-})=E_{X}-E_{X^{-}}$ were found to be $E_{B}(X)=-1.3\: meV$ and $E_{B}(X^{-})=+6.3\: meV$ respectively. From $0-0.4V$, Using bias, we can change the laser detuning $\Delta_{L}$ in resonance fluorescence, by varying the transition energy $E_{X}$ with respect to a fixed continuous wave laser. This is a powerful technique because it enables ultrahigh frequency variation of laser detuning, as we will show later. We also gain the freedom to study both the photoluminescence and resonance fluorescence of more than one exciton species with a constant laser energy $E_{L}$. Distributed Bragg reflectors enclosed the device with seven repeats below and four on top, providing enhanced excitation and collection efficiency at $E=1.324\:eV$. The sample was mounted at $45^{\circ}$ to the laser polarization, and maintained at $10\:K$ using a dewar-insert cryostat. Single quantum dots were optically addressed via a microscope setup in a confocal arrangement. Laser rejection was achieved by crossing the polarization states of detection and excitation photons \cite{NickVamivakas2009,Kuhlmann2013}, with extinction ratios of upto $10^{9}$ and fluorescence to laser scatter ratio of exceeding $200:1$. The collected signal was resolved using a spectrometer and a charge coupled detector. Photon statistics were analysed using a Hanbury-Brown and Twiss apparatus with two avalanche photodiodes (APD) and timing electronics.

\begin{figure}[]
\includegraphics[width = 6cm]{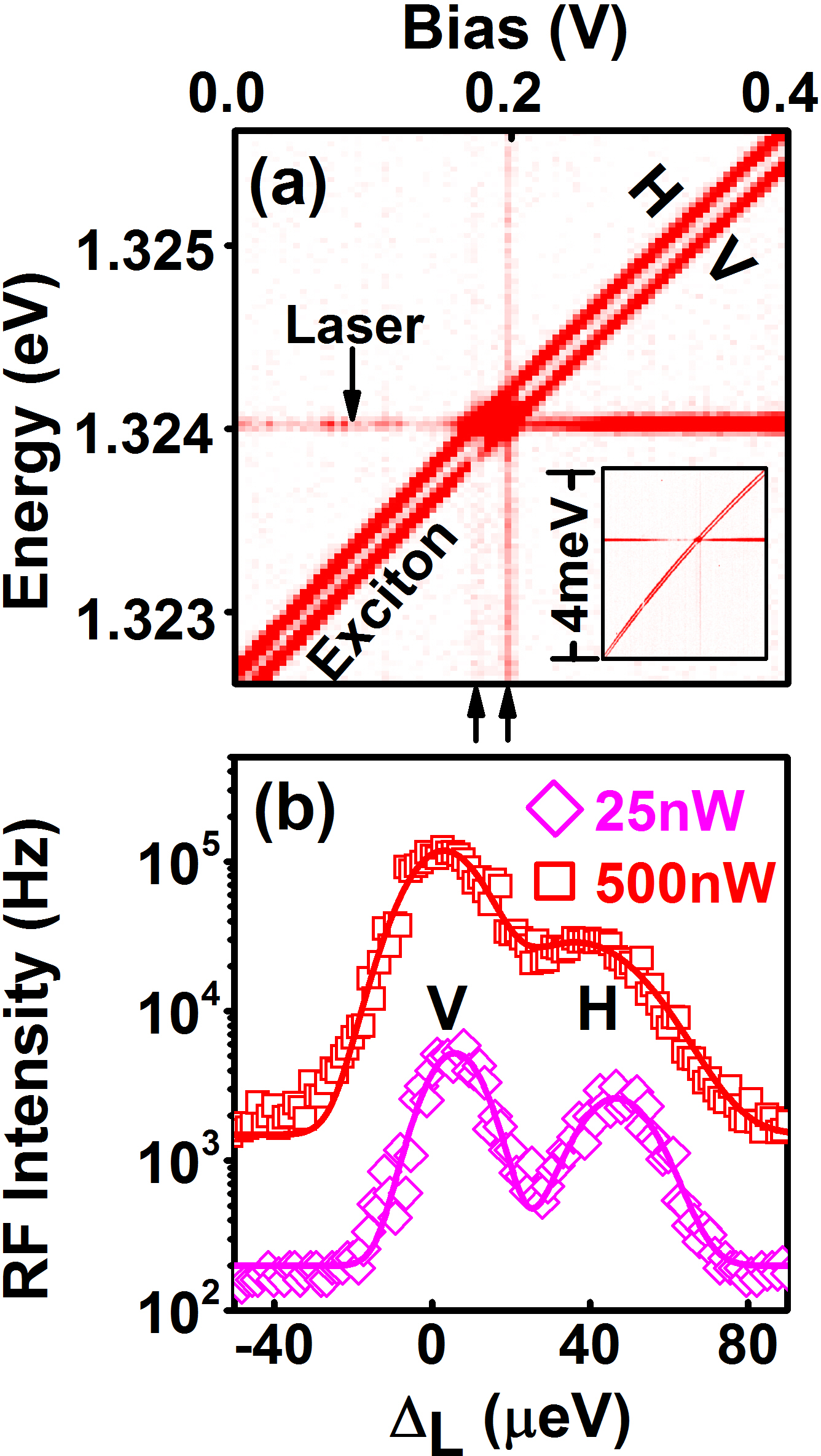}
\caption{Fluorescence spectroscopy of a single quantum dot. (a): Spectrally resolved exciton doublet tuned into resonance with laser energy fixed at $1.324\:eV$.  Inset: Full range of detuned emission. (b): Fluorescence measured by an APD as the exciton resonance was tuned through the laser energy.}
\label{Fig1}
\end{figure}

Figure \ref{Fig1}(a) shows how the emission energies of horizontally (H) and vertically (V) polarized eigenstates of an exciton were electrically tuned through the resonance with a fixed energy laser at $E_{L}=1.324\:eV$, at $\sim0.2V$. the intensity of the back-scattered laser varies by an order of magnitude whereas collected signal from the exciton emission remains approximately constant, for nonzero $\Delta_{L}$. The variation in laser intensity (the horizontal line in Figure \ref{Fig1}(a)) can be explained by a bias-dependent birefringence in the sample, changing the conditions for optimal laser rejection. The inset figure shows the full energy range in this measurement, it shows that emission from the exciton persists even when the laser was several $meV$ away from resonance. A recent study \cite{Weiler2012} also observed this, and attributed it to the interaction between the exction and a bath of longitudinal acoustic phonons with $\sim5\:meV$ bandwidth. The off-resonant emission occurs by absorbing ($\Delta_{L}<0$) or emitting ($\Delta_{L}>0$) phonons, where $\Delta_{L}=E_{X}-E_{L}$ is the laser detuning. Figure \ref{Fig1}(a) shows energetically broad features in the emission spectrum that appear as vertical lines, at biases corresponding to resonances of each exciton eigenstate (labelled with double arrows). We believe that the broad features we have observed in \ref{Fig1}(a) derive from phonon scattering.

We now focus on the resonance fluorescence in the bias range close to the resonance at $0.2V$, the spectrum was measured in fine steps for high ($500\:nW$) and low ($25\:nW$) excitation laser powers. In Figure \ref{Fig1}(b), the respective linewidths at high excitation power were, $\Gamma_{H}=36.8\pm1.6\:\mu eV$ and $\Gamma_{V}=21.3\pm0.6\:\mu eV$, respectively. The difference in intensity and linewidth between components of the doublet can be accounted for by the transition eigenstates not being aligned at $45^{\circ}$ to the excitation and detection axes. The fact the transition with greater apparent width has lower maximum intensity supports this claim. These linewidths are significantly broader than those observed in similar studies \cite{Nguyen2011,Matthiesen2012}. We attribute the broadening to an ensemble of fluctuating charges that maybe located within nearby impurities and the superlattice or the contact layers \cite{Kuhlmann2013}. These fluctuations change the transition energy through the Coulomb effect. The detuning spectrum, shown in Figure \ref{Fig1}(b) then represents a Gaussian probability distribution of the ``noise'' in $E_{X}$. Power broadening \cite{Citron1977} was also observed, yet even at the lowest excitation powers, linewidths were still broad with $\Gamma_{H}=21.7\pm1.2\:\mu eV$ and $\Gamma_{V}=16.7\pm0.5\:\mu eV$.

\begin{figure}[]
\centering
\includegraphics[width=6cm]{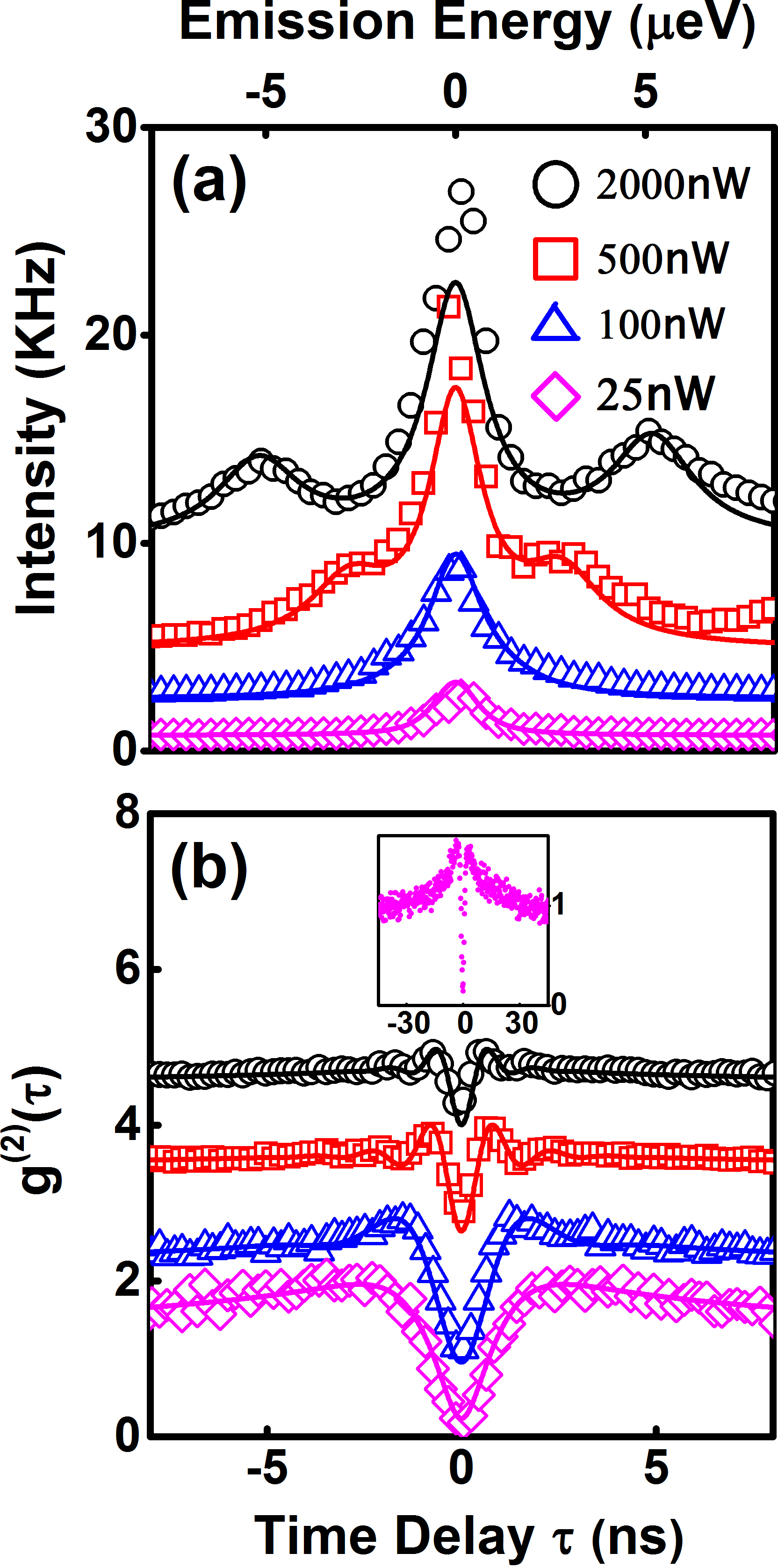}
\caption{Excitation-power dependent resonance fluorescence. In all the plots, data ($\circ$) are superimposed by theoretical simulations (solid line) with experimentally obtained parameters. (a): Resonance fluorescence spectrum for different excitation powers. (b): $g^{(2)}(\tau)$ recorded at various excitation powers, showing the Rabi oscillations; inset: $g^{(2)}(\tau)$ recorded at $25\:nW$ excitation power, showing the full range of bunching over $40\:ns$.}
\label{Fig2}
\end{figure}

We determined the coherence time $T_{2}$ of the photon by measuring the spectrum of resonance fluorescence. Figure \ref{Fig2}(a) presents the resonance fluorescence spectrum of one transition of the doublet in the exciton's fine structure, for a range of laser powers. This measurement was recorded using a Fabry-P\'{e}rot interferometer of $0.8\mu eV$ bandwidth and an APD. The transformation of the spectrum with respect to power correlates to a competition between the power dependent coupling strength $\Omega$, and the rate of decoherence $\Gamma_{s}$ in the system. The magnitude of $\Omega$ bears a direct indication of the amount of energy exchange between laser photons and the exciton. For lower powers the coupling is weak $\Omega\ll\Gamma_{s}$ and Rabi effects are small. At $25\:nW$ the linewidth was measured to be $\Gamma=1.3\pm0.1\:\mu eV$. This corresponds to a photon coherence time of $T_{2}=1.0\pm0.1\:ns$. Measuring the spectrum on resonance effectively filters out the ``noise'' in $E_{X}$, as absorption only takes place when the emission energy is resonant with the laser. At higher powers, the laser-exciton coupling becomes strong enough to create a dressed-ladder system \cite{Cohen-Tannoudji2008}, in which additional optical paths splits the central (Rayleigh) peak with energy $E_{X}$, to red and blue shifted sidebands with energies $\pm\hbar\Omega$. In the saturation limit, $\Omega\gg\Gamma_{s}$, Rabi effects are significant and multiple Rabi oscillations may occur before radiative recombination.

Theoretical simulations of the resonance fluorescence spectrum were produced based on a polaron master equation model, which assumed a weakly coupled phonon bath \cite{McCutcheon2013}. At resonance, the spectrum has the following analytical form:

\footnotesize
\begin{eqnarray}
  S(\omega)=\frac{\Gamma_{2}^{2}}{2(\Gamma_{2}^{2}+\omega)^{2}}
  +K\left(\frac{\Gamma_{s}}{\Gamma_{s}^{2}+(\omega+\Omega_{r})^{2}}+
  \frac{\Gamma_{s}}{\Gamma_{s}^{2}+(\omega-\Omega_{r})^{2}}\right),\label{IncohSpect}
\end{eqnarray}

\normalsize
where $\Gamma_{s}=(\Gamma_{1}+\Gamma_{2})/2$, $\Gamma_{1,2}$ are the recombination and pure dephasing rates respective to the numeric labels. $\Omega_{r}=\sqrt{\Omega^{2}+\epsilon^{2}-\Gamma_{s}^{2}}$ is the Rabi frequency corrected for decoherence due to reservoir of phonons, by the $\epsilon^{2}$ term, as well as the dephasing, by $\Gamma_{s}$. The constant $K$ was defined as a function of $\Gamma_{1,2}$ and $\Omega_{r}$. We take the experimentally determined lifetime broadening and pure dephasing rates of $1\:GHz$ respectively in the calculation. We have extended the model to take into account of the broadening, observed in Figure \ref{Fig1}(b), by weighting the resonance fluorescence spectrum with a Gaussian distribution $16.7\:\mu eV$ wide, which described the spectral noise. At the highest powers a discrepancy between the model and the data at zero detuning can be explained by an incomplete suppression of laser scatter.

Antibunching statistics of the resonant fluorescence were explored by measuring the second order intensity correlation for the same excitation powers, as presented in Figure \ref{Fig2}(b). At low powers the observed strong antibunching with $g^{(2)}(0)=0.040\pm0.005$ implies emission from a single quantum state. At higher powers strongly damped oscillations symmetric about zero time delay are indicative of Rabi oscillations. At $500\:nW$ the Rabi oscillation period was found to be $T_{\Omega}=1.6\pm0.2\:ns$, consistent with a sideband splitting of $2.6\:\mu eV$, measured in Figure \ref{Fig2}(a). Extracted Rabi frequencies for each power were used to theoretically simulate $g^{(2)}(\tau)$ using \cite{Scully1997},
\begin{eqnarray}
    g^{(2)}(|\tau|)=1-e^{-\beta|\tau|}
    \left(cos\theta+\frac{4\Gamma_{s}}{\sqrt{\Omega^{2}+4\Gamma_{s}^{2}}}
    sin\theta\right) \label{g2Function},
\end{eqnarray}
where $\beta$ is a constant defined in terms of $\Omega$ and $ \Gamma_{s}$ and $\theta=\theta(|\tau|)$. As power is increased, the oscillatory frequency increases as the square root of power and the width of the antibunching ``dip'' local to zero time narrows. The antibunching time is limited by $\Gamma_{s}$, and also the detector resolution of $480\:ps$. A long-time bunching effect, shown by the inset of Figure \ref{Fig2}(b), also exists in the data, which decays on the duration of $40\:ns$ at the lowest power, extending to nearly $100\:ns$ at the highest power. Spectral fluctuations in the transition energy can cause this bunching, and we hypothesize that it is caused by the same mechanism that leads to the broadening seen in Figure \ref{Fig1}(b). However, the power-dependence is not well understood at this time. Convolution of equation 2 with a double sided exponential decay describing the bunching, as well as the measured instrumental response gave a near perfect fit to the data points. In spite of spectral fluctuations and broadening caused by the charge noise, we observed Rabi oscillations and strong antibunching signatures under continuous wave excitation, which implies that the resonant driving field can still generate pure photon states.

\begin{figure}[]
\centering
\includegraphics[width=8.5cm]{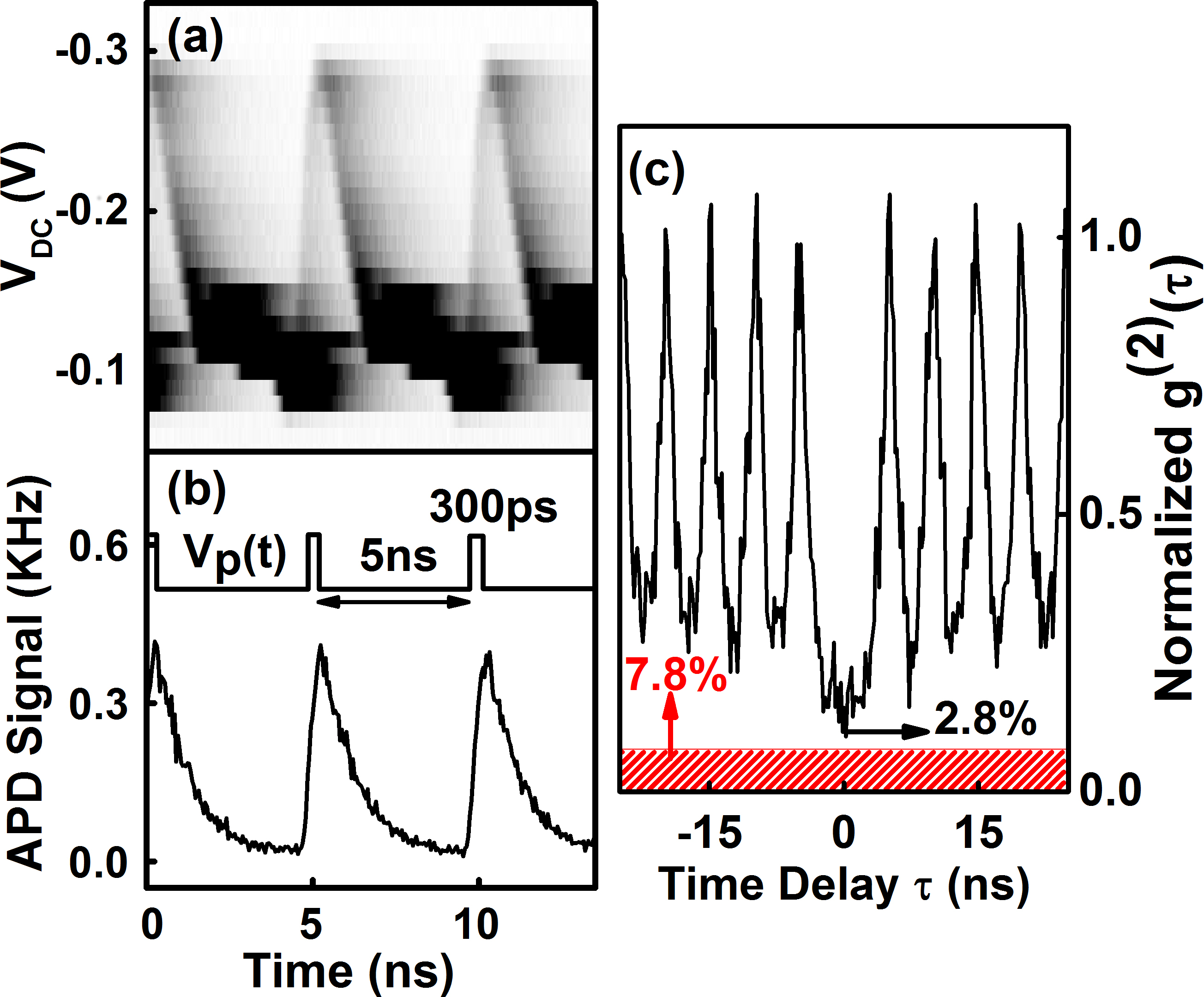}
\caption{Resonance fluorescence of the QD with electrical modulation, driven by a continuous wave laser. (a): Time-resolved emission spectra of the modulated emission recorded as the constant DC offset is varied. (b): Time-resolved emission spectrum at $V_{DC}=-0.29\:V$. Drawing shows the square waveform pulses $V_{p}(t)$. (c): Second-order intensity correlation measurement at $V_{DC}=-0.29\:V$, $200\:$\textit{MHz} pulses and $300\:ps$ pulse width.}
\label{Fig3}
\end{figure}

A key requirement for quantum communication is to deterministically produce single photons on demand. Given that our sample design enables the manipulation of exciton energy, we are able to generate a pulsed stream of single photons from the CW laser, by applying a rapidly oscillating bias $V_{P}(t)$ to the device, as shown by the drawing in Figure \ref{Fig3}(b). $V_{P}(t)$ consists of a sequence of alternating current pulses superimposed on a constant offset voltage $V_{DC}$. During one pulse cycle, an initially off-resonant exciton eigenstate is rapidly Stark shifted into resonance during pulse rise, remains there for the duration of pulse width and finally Stark shifted out of resonance during pulse fall. This process repeats itself for each subsequent pulse in $V_{p}(t)$. Each time a pulse is applied to the quantum dot, an exciton state is prepared resonantly, triggering single photon emission via spontaneous decay.

We assessed the performance of the electrical trigger by studying the time resolved fluorescence spectra of the exciton as a function of $V_{DC}$, for a range of different pulse parameters. Figure \ref{Fig3}(a) shows typical time resolved spectra recorded for varying $V_{DC}$. At $V_{DC}=-0.29\:V$, the exciton resonance intersects the laser only once during a single pulse period. For efficient single photon emission, well isolated pulses are required in addition to a faster triggering time, compared to the transition's radiative lifetime. The time resolved fluorescence spectrum recorded at $V_{DC}=-0.29\:V$ and $300\:ps$ pulse width is shown in Figure \ref{Fig3}(b), from which we extracted the lifetime to be $T_{1}=1.0\:ns$.

The pulsed source was characterized by measuring the second order intensity correlation function $g^{(2)}(\tau)$, using a nonpolarizing beam splitter and two APDs. A perfect single photon source has $g^{(2)}(0)=0$ while a Poissonian source has $g^{(2)}=1$. Figure \ref{Fig3}(c) shows the normalized intensity correlation function recorded for electrically triggered fluorescence photons with $V_{DC}=-0.29\:V$, repetition rate $200\:$\textit{MHz}. Background laser scatter contributed $\sim7.8\:\%$ of the total detected correlation signal. With background correction, we observe that $g^{(2)}(0)=0.028\pm0.010$, which corresponds to the significant suppression of the zero-time peak, and hence single photon emission dominates.

We have demonstrated the generation of strongly antibunched single photons from a quantum dot, using a hybrid scheme that combines both electrical and resonant optical control. The conducted studies on resonance fluorescence were in the incoherent regime where there is a finite population of the upper state and the fluorescent photons are not phase coherent with the laser. To further explore coherent dynamics of the ultrafast single photons source, we may consider the ultra-low power Heitler regime \cite{Nguyen2011,Matthiesen2012}.

\begin{acknowledgements}
We thank A. Nazir, J.I. Smith and T. Rudolph for fruitful discussion of resonance fluorescence modelling. Y.C. gratefully acknowledges financial support from both EPSRC CDT in Controlled Quantum Dynamics at Imperial College London and Toshiba Research Europe Ltd in Cambridge.
\end{acknowledgements}


\end{document}